\documentclass{webofc}
\usepackage{hyperref}
\usepackage[varg]{txfonts}   
\begin{document}
\title{Particle production and entropy measurement in ALICE}
%
%

\author{\firstname{Alek} \lastname{Hutson}\inst{1}\thanks{\email{alek.hutson@cern.ch}} \fnsep On behalf of the ALICE collaboration
}

\institute{University of Houston 
\
          }

\abstract{%
  One of the main goals in the study of hadronic interactions at LHC energies is the attempt to characterize the mechanisms involved in particle production in different regimes. The charged-particle multiplicity is one of the most interesting observables in these kind of studies. Measurements of charged-particle pseudorapidity densities in pp collisions at $\sqrt{\textit{s}}$ = 13.6 TeV are presented, for the first time. Furthermore, we present a method for better understanding the collision dynamics. On the one hand, the pseudorapidity dependence of charged-particle production provides information on the partonic structure of the colliding hadrons and is sensitive to non-linear QCD evolution in the initial state. On the other hand to understand the thermal-like behavior and particle yields in pp collisions, a possible approach is to address the principles of quantum states and their entanglement in the produced system. The entanglement in the initial state has a measurable effect on the evolution of the system and is the driving mechanism behind the thermal-like behavior and particle yields observed. We describe a method to understand the level of entanglement in the initial and final states of the collision using a calculation of entropy, the final-state entropy being calculated from multiplicity distributions.
}
\maketitle
\section{ALICE Run 3 upgrade}
\label{intro}
Aside from the higher energies reached in the LHC's Run 3, the ALICE Run 3 upgrade marks a significant enhancement in detector performance. Advancements in three critical systems used in multiplicity measurements were made: the Inner Tracking System (ITS), the Time Projection Chamber (TPC), and the Fast Interaction Trigger (FIT). 

Hardware upgrades have been made to the ITS, which now features seven layers of highly granulated silicon pixel detectors, improving the track pointing resolution, readout rate, and tracking efficiency, particularly for low-\textit{$p_t$} particles, and boosts the readout rate capabilities up to to 1 MHz, allowing for more precise and rapid data collection. 

The TPC, serving as the primary subsystem for particle tracking in the central barrel, has seen its Multi-Wire Proportional Chambers (MWPC) replaced with Gas Electron Multipliers (GEM), which alleviates rate limitations and minimizes space-charge distortions—critical for enhancing the accuracy of multiplicity measurements. 

Finally, the FIT has been augmented with additional Cherenkov modules, increasing the coverage in pseudorapidity and providing a time resolution of approximately 33 ps. These improvements collectively contribute to a more refined measurement of multiplicity, ensuring that ALICE's capabilities in high-precision tracking and detection remain at the forefront of particle collision research.

\section{Multiplicity Measurements}
\label{sec-1}
Multiplicity measurements are made by counting tracks coming from the primary vertex. In any collision experiment certain quality cuts are made at both the event level and track level. At the event level the FIT is used as our min. bias event trigger, ensuring the timing of collisions coincides with filled bunch crossing. Furthermore, the ITS is used to reconstruct the primary vertex and ensure collisions are near the center of the symmetric tracking systems. The ITS and TPC are the subsystems used for track counting. To ensure the quality of our tracks we require at least one hit in the ITS that matches to a track reconstructed within the TPC. Many more standard quality cuts were made to ensure good data quality. 

Correction for detector effects must also be made to ensure an accurate measurement. The correction process involves a Bayesian unfolding in which Monte-Carlo simulations are used to simulate events and detector response. Details regarding typical cuts and corrections can be found in Ref.\cite{alice}.

Multiplicity distributions and densities are given in three different event classes: Inelastic (INEL), inelastic with at least one particle in $|\eta|<0.5$ (INEL$>$0), and non single diffractive (NSD).
Previous measurements of multiplicity density have shown a relatively flat distribution across $\eta$ for all event classes \cite{alice}. Also, the mean multiplicity at $\eta$ = 0 has followed a power law trend as a function of collision energy.

In Fig.1 preliminary multiplicity density results are shown for pp collisions at center of mass energy $\sqrt{\textit{s}}$ = 13.6 TeV for the INEL$>$0 event class at mid-rapidity.

\begin{figure}[h]
\centering
\sidecaption
\includegraphics[width=.6\textwidth]{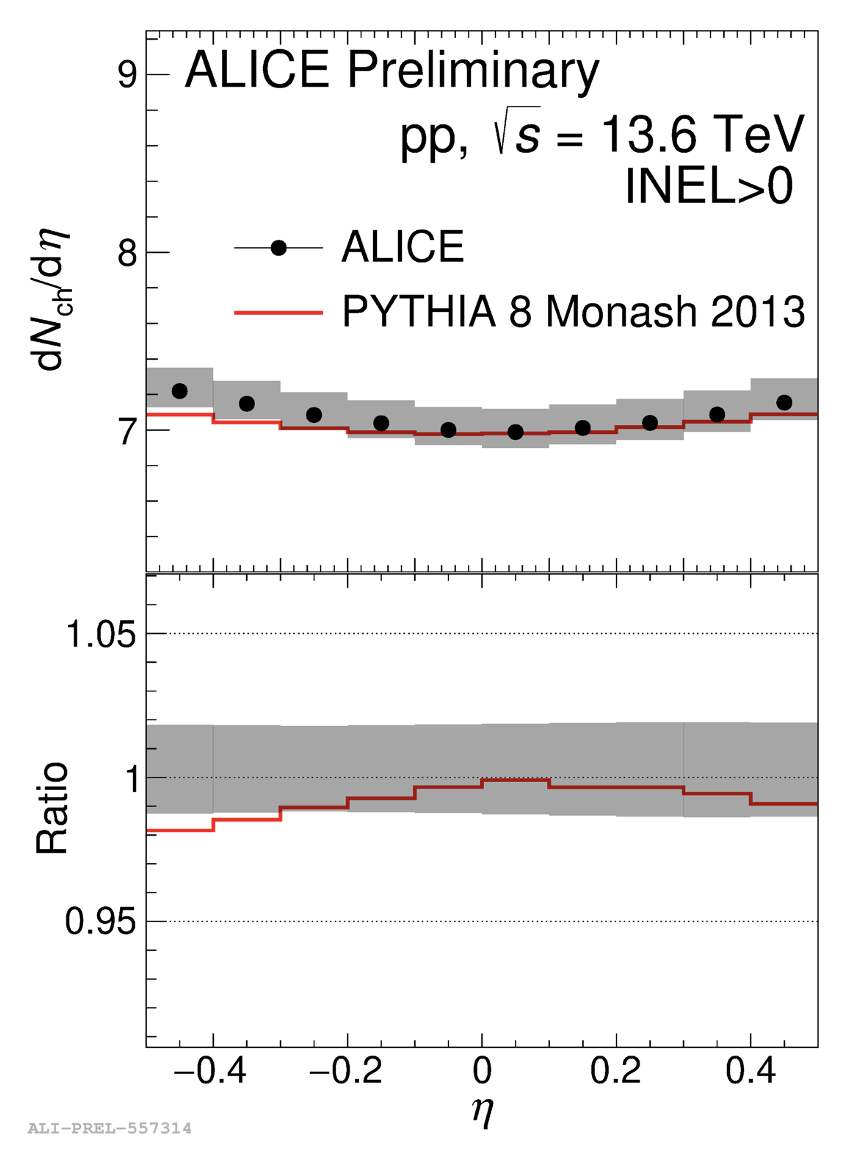}
\caption{Multiplicity density for INEL$>$0 multiplicity class at $\sqrt{\textit{s}}$ = 13.6 TeV, in comparison to the multiplicity predicted by event generator PYTHIA 8.}
\label{fig-1}       
\end{figure}

\begin{figure}[h]
\centering
\includegraphics[width=.8\textwidth]{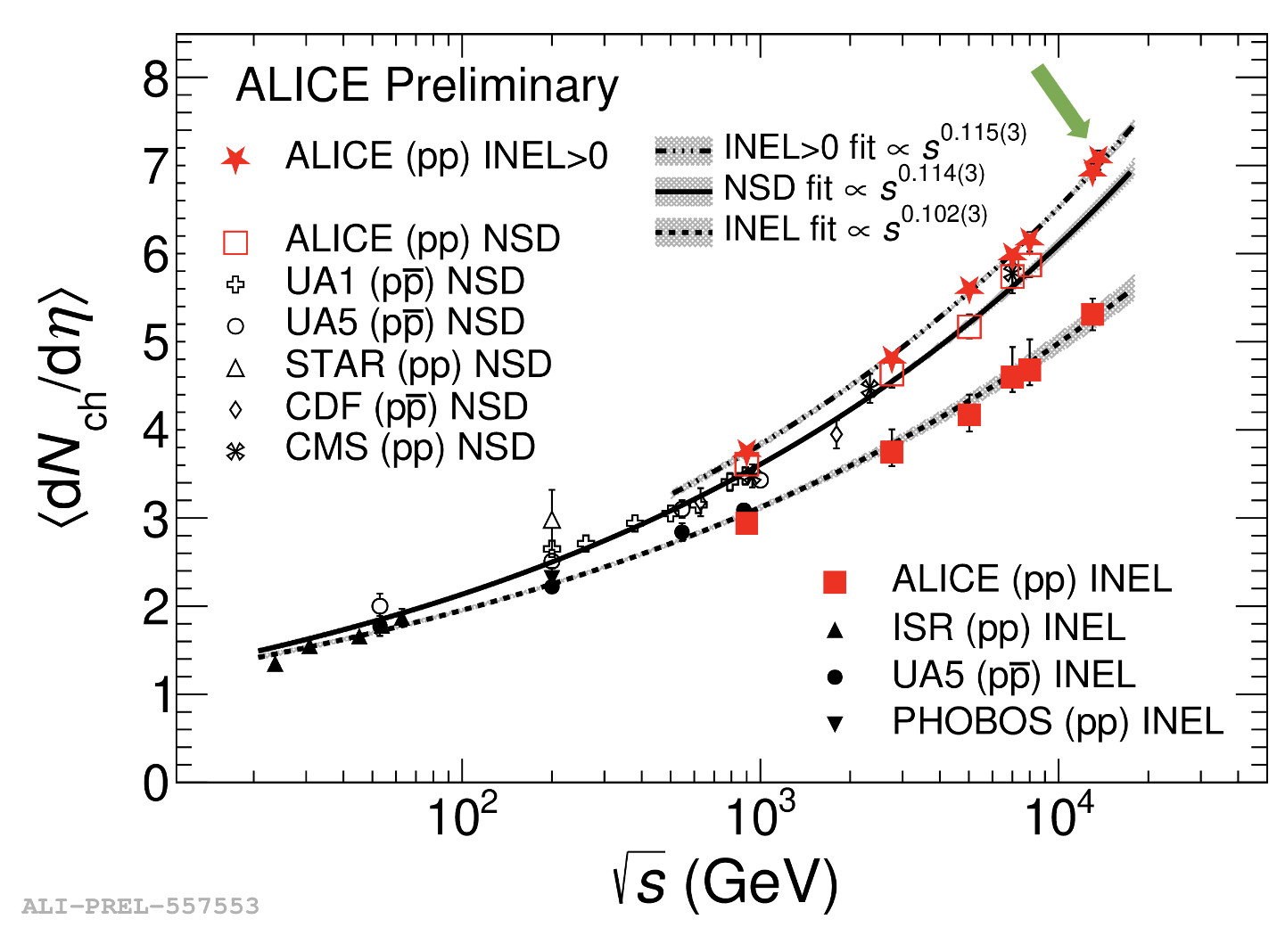}
\caption{ Average multiplicity densities at $|\eta|=0$ as a function of collision energy.}
\label{fig-2}       
\end{figure}
\newpage
Fig.2 shows the average multiplicity density as a function of collision energy, including the new result at 13.6 TeV.

\section{Entanglement Entropy}
\label{sec-2}
It is our goal to better understand the thermal-like behavior and particle yields observed in pp collisions using fundamental quantum mechanics and employing the formalism of quantum information. We seek to establish an equivalence between the initial-state entropy, calculated using Parton Distribution Functions (PDFs), and the final-state entropy derived from multiplicity distributions of primary charged particles, as measured by the ALICE detector. This equivalency has been seen in e-p interactions \cite{hentschinski2022}; however, complications arise in a system where there are two distributions of partons. The formalism for making this comparison in pp collisions was established in part in Ref.\cite{tu2019}. 

\subsection{Final-State}
\label{sec-2}
In the final-state we can calculate a thermodynamic entropy using the probability distribution of produced particles. By setting pseudorapidity boundaries and counting particles in a range we can make a meaningful comparison to the initial-state integrated over the same kinematic range. This range can be any width in $\eta$ as long as it corresponds to the same kinematic range integrated over when defining the initial state. For this analysis it is convenient to measure charged particles at mid-rapidity where we have strong particle tracking abilities and high statistics. 

\subsection{Initial-State}
\label{sec-2}
In pp collisions, entropy in the initial system originates from a break in entanglement between a probed and untouched region. By tracing over the degrees of freedom within the reduced density function of the system one can calculate an entanglement entropy between these regions. To simplify this complex task due to the diverse internal structure of protons, the area of interest is restricted to regions dominated by indistinguishable gluons such that the internal degrees of freedom are simplified to a number of partons. For small momentum fractions \textit{x}, the initial entropy reduces to the logarithm of the number of partons over a given \textit{x} range. This is determined by integrating over the known parton probability distribution, with integration limits based on the spatial region measured in the final-state.

\section{Conclusions}
\label{sec-3}
Multiplicity measurements were made for INEL$>$0 event class at $\sqrt{\textit{s}}$ = 13.6 TeV at mid rapidity. These preliminary results show good agreement with multiplicity estimations made using PYTHIA 8 Monash 2013, demonstrating that event generators continue to accurately reproduce particle multiplicities at higher energy. Furthermore, the mean multiplicity at  $\eta$ = 0 follows the expected power law trend as a function of collision energy.

In future studies we hope to use these measurements to show equivalency between the entropy calculated from hadron multiplicity in the final state and parton multiplicity in the initial state, in the low-\textit{x} kinematic regime.

%
%
%

\end{document}